# The ALTO Project at IPN Orsay

*F. Ibrahim*

CNRS/IN2P3/Université Paris Sud
*Institut de Physique Nucléaire d'Orsay, F-91 406, Orsay*

*Cedex, France*

E-mail address: ibrahim@ipno.in2p3.fr

**Abstract** :

In order to probe neutron rich radioactive noble gases produced by photo-fission, a PARRNe1 experiment (**P**roduction d'**A**tomes **R**adioactifs **R**iches en **Ne**utrons) has been carried out at CERN. The incident electron beam of 50 MeV was delivered by the LIL machine: **L**EP **I**njector **L**inac. The experiment allowed to compare under the same conditions two production methods of radioactive noble gases: fission induced by fast neutrons and photo-fission. The obtained results show that the use of the electrons is a promising mode to get intense neutron rich ion beams. Thereafter, the success of this photo-fission experiment, a conceptual design for the installation at IPN Orsay of a 50 MeV electron accelerator close to the PARRNe-2 device has been worked out: ALTO Project. This work has started within a collaboration between IPNO, LAL and CERN groups.

## 1. Introduction:

There is currently in the nuclear physics community a strong interest in the use of beams of accelerated radioactive ions. Although a fast glance at the nuclide chart immediately shows the vast unknown territories on the neutron-rich side of the valley of beta stability (see figure 1), only few projects are concerned with the neutron-rich nuclides.

The availability of intense n-rich ion-beams will open new perspectives in the study of nuclei very far away from the valley of stability. It would allow to apprehend the behaviour of the nuclear matter under extreme conditions [1,2]. Several laboratories have concentrated their efforts in studies aiming to produce beams intense enough for the next generation of experiments (SPIRAL II and EURISOL projects). To get such beams, a large R&D effort is required. Uranium fission is a very powerful mechanism to produce such radioactive beams.

A substantial part of the PARRNe program (**P**roduction d'**A**tomes **R**adioactifs **R**iches en **Ne**utrons) at the IPN Orsay is dedicated to the development of n-rich isotope beams by the ISOL method.

## 2. Investigation of fast neutrons production mode

The aim of the PARRNe program is to establish the feasibility of producing neutron-rich radioactive beams for the SPIRAL-II project at GANIL, and to determine the optimum conditions for the production of such beams. The neutron-rich radioactive nuclides are to be produced by fissioning a heavy nuclide, such as $^{238}$U. The technique originally proposed [3] is the use of energetic neutrons to induce fission of depleted uranium. The neutrons are generated by the break-up of deuterons in a thick target, the so called converter, of sufficient thickness to prevent charged-particles to escape. The energetic forward-going neutrons impinge on a thick production target of fissionable material. The resulting fission products accumulate in the target, diffuse to the surface from which they evaporate, are ionised, mass-selected and finally post-accelerated. An ISOL (Isotopes Separator On- Line) device has been developed to carry out various R&D works [4] (see figure 2).

This method has several advantages. The highly activated converter can be kept at low temperature without affecting the neutron flux. The target is bombarded by neutral projectiles losing energy only by useful nuclear interactions and having a high penetrating power allowing very thick targets.

One of the main objectives of the R&D program was to determine the energy of the primary deuteron beam giving the best yields of radioactive nuclides of interest for radioactive beams while taking into account beam power evacuation and safe operation of the facility. The approach has consisted in carrying out simulations with various codes available or developed by different task groups of the SPIRAL-II project and performing a number of key experiments to validate the simulations. In this way, confidence is gained about the predictive power of the codes for situations where experiments could not be set up within the allocated time for the study.

The concept of using neutrons generated by deuteron break-up implies a study of production yields, energy spectrum and angular distributions of neutrons in converters made of various materials and as a function of deuteron energy. Experiments were performed at IPN-Orsay,

KVI-Groningen and Saturne at Saclay. They explored a range between 14 and 200 MeV deuteron energy. The main features of neutron spectra are listed below

At forward angles, the energy distribution has a broad peak centred at about 0.4 times the deuteron energy. The angle of emission becomes narrower with increasing energy. For 100 MeV deuterons, the energy width (FWHM) of the neutron spectrum is about 30 MeV and the FWHM opening angle of the cone of emission is about 10 degrees [5,6].

There is a rather isotropic distribution of neutrons of a few MeV due to evaporation in fusion reaction.

The angular distributions and energy spectra are in fair agreement with calculations with an extended version of the Serber model [7] and with the LAHET code. The Serber model reproduces the distributions of high-energy neutrons but not of the low-energy neutrons since evaporation is not implemented in the code. LAHET reproduces the low energy neutron spectrum while it tends to slightly underestimate (less than a factor 2) the neutron distributions at very forward angles.

A strong increase in neutron production is observed between 14 and 100 MeV deuteron energy. It is much less pronounced between 100 and 200 MeV (see figure 3). Among converters tested Be is slightly more productive than C. It has, however, disadvantages related to its physical and chemical properties.

Productions of radioactive noble gases on a cryogenic finger have been measured for different deuteron energy in the same experimental conditions with the so called PARRNe 1 set-up [8] in the framework of the European RTD program SPIRAL-II. The PARRNe1 set-up has been designed compact and portable to enable its installation at various accelerators. The search of the optimal energy of the deuterons was done by installing this set-up successively at IPN Orsay (20 MeV) [9], at CRC Louvain La Neuve (50 MeV) and at KVI Groningen (80 and 130 MeV) [10].

## 2. Investigation of Photo-fission production mode

It has recently appeared that photofission could be an alternative to n-induced fission [11,12]. We have therefore initiated a study of photo-fission induced by Bremsstrahlung generated by electrons.

With an electron driver, electron interaction with matter will radiate Bremsstrahlung photons inside the target. Fission will then be induced by those photons exciting the Giant Dipolar Resonance (GDR) of the nucleus at the right energy. This well-known process is called photofission.

The GDR cross section for $^{238}$U is shown in the figure 4. A maximum fission probability of 160 mb is obtained for photons having energy around 15 MeV. At that energy, the photoelectric and the Compton and Rayleigh scattering cross sections are starting to fall off rapidly so the main contributions to gamma absorption are $e^+e^-$ pair production and the photonuclear reactions (γ,f), (γ,n) and (γ,2n). Although the absolute fission cross section is rather small (compared to normal fission with neutrons), its contribution is not negligible as even a pair production reaction may in a thick target eventually lead to a fission through the resulting photon produced. In the same manner the neutrons produced by (γ,n) and (γ,2n) reactions as well as the (γ,f) itself can also induce fission, this time by the regular (n,f) high cross section (0.5 barn for fast neutrons). Therefore, in a thick target, photofission may be a rather interesting way of creating radioactive fission fragments.

Unfortunately, no efficient monochromatic sources of 15 MeV photons are available. The most common way for producing high gamma fluxes is the Bremsstrahlung radiated by passage of electrons through matter. This process has a cross section rising linearly with energy. It will dominate the ionization process above a critical energy (around 20 MeV). But the resulting Bremsstrahlung spectrum is widely spread in energy from zero up to the full initial electron energy (figure 4). Although each single electron may ultimately produce as high as 20 photons, only a small fraction of it (0.5 to 0.7 gamma per e$^-$) are "useful" photons lying in the GDR range (15 ±5MeV).

A simple calculation including the main electron interactions (Bremsstrahlung and ionization) and the main nuclear reactions of interest (pair production and fission cross section) in a thick depleted uranium target can give the expected number of fission per incident electron. In figure 5, the number of fissions produced by the (γ,f) reaction is plotted as a function of the electron energy [12]. This result is a complete Monte Carlo calculation performed with a MCNP code offering also photonuclear capability (full electron, photon and neutron transport). The obtained result is in accordance with the simple analytic calculations. For comparison, fission production is also given when using a tungsten converter (5 mm thick) in front of the $^{238}$U target. It appears that when using a converter in the electron driver option, less than 30% of the beam power is lost inside the converter (in contrast to the d driver option). In the direct method one will produce about 25% more fission per electron (and probably more when taking in account neutron induced fission). Fission production is almost linear above a threshold energy of 10 MeV. High production is obtained above 40 MeV.

In order to compare rapid neutron induced fission and photofission, measurements of Kr and Xe isotopic distributions produced by photofission and diffused out of a thick UCx target has been performed using the same PARRNe1 device in the same conditions that with deuterons beams and with the same target [13]. The set-up of the experiment is presented in figure 6.

The incident electron beam of 50 MeV was delivered by the LIL machine: *LEP Injector Linac*. The electrons are slowed down in a W converter or directly in the target, generating Bremsstrahlung γ-rays which may induce fission.

The PARRNe-1 set up consists in measuring the activity of produced radioactive noble gases by trapping them on a cold finger (13°K) in front of which is placed a Germanium detector. The cold finger is connected to the target by an 8 meter long tube at room temperature. This device allows to shield the detection system from the irradiation point. All other produced elements are condensed at the entrance of the long tube.

The measurements have been done with a 4 mm W converter in different position (8 mm from the target, 4 mm from the target) and one measurement without W converter. Comparison with the 80 MeV deuteron induced fission measurements are presented in figure 7. The results obtained are well understood taking into account the percentage of photons between 11 and 17 MeV emitted in the cone subtended by the target that is the solid angle [13].

The extrapolation of the obtained results indicates that the use of a 50 MeV electron beam of 10 µA, would allow a gain in production of at least 100 in comparison with PARRNe-2 results (using 26 MeV deuteron beam of 1 µA). Such a situation would offer the possibility to provide for example **~4x10$^7$** $^{132}$Sn /s and **~2x 10$^5$** $^{78}$Zn /s with the PARRNe 2 device at Orsay.

## 3. The ALTO project

It has been decided to start a conceptual project for the installation at IPN-Orsay of 50 MeV electron accelerator [14] : The ALTO project (**A**ccélérateur **L**inéaire auprès du **T**andem

d'**O**rsay). The accelerator will be installed in the experimental area of the Tandem, to deliver beams maily to PARRNe-2 device. After the photo-fission experiment success, the CERN scientists authorities interested in the ALTO project have decided to offer the LIL front end to the IPN Orsay.

The linac is composed of thermionic gun, a bunching system and a matching section to the linac. The schematic lay-out in the figure 8, presents an overall view of the main components of ALTO.

**The gun** is a thermionic source held to 90 kV, it is designed to provide beam pulses up to ~ 2 µsec in length and peak current of 50 mA. The operating frequency is 100 Hz.

**The pre-buncher** is a RF cavity 15 mm long working in standing waves mode at 3 GHz. It is mounted at 100 mm from the buncher. For the transverse focusing of the beam, three solenoids are installed downstream the gun.

**The buncher** is a tri-periodical RF structure cavity working in standing waves mode at 3 GHz. The cavity structure is surrounded by a solenoid producing a 0.2 T magnetic field. The buncher provides an output energy of about 4 MeV.

**The accelerating section** is a 4.5 m long RF cavity operating in travelling waves mode at 3 GHz. The section output energy is of 46 MeV. In order to match the beam from the buncher exit to the accelerating section entrance we use one solenoid and one quadrupole triplet.

The whole RF structure (pre-buncher, buncher and accelerating section) is powered by only one 35 MW klystron TH 2100. The operating of the accelerator at 50 MeV needs a HF power less than 20 MW.

The transport beam line consists of two 65° dipole magnets (R=0.4 m) and seven magnetic quadrupoles. The first Q-triplet placed behind the accelerating section allows the control of the beam envelop at the entrance of the first magnet. To make the achromatism in the deviation, a quadrupole will be placed between the two magnets. The spot beam dimension is adjusted on the PARRNe target by using the last Q-tripltet. The expected energy resolution is less than $5 \times 10^{-3}$. The beam line is equipped by instruments for the beam diagnostic: measurement of current, beam position, energy and energy dispersion.

The installation cost of ALTO is estimated to 1.4 M€ without including the cost of the LIL front end (from CERN), the HF material (from LAL), and all the infrastructure and manpower (from IPNO). After acceptation of the financial plan before the end of 2002, the ALTO project planning will be extended within a maximal period of 18 months.

The expected intensities for the 30 keV mass separated beams at ALTO will give the opportunity for nuclear spectroscopy of very neutron-rich nuclei in various region of interest.

Properties such as the modification of shell closures are predicted for a wide range of extremely neutron-rich nuclei. Nuclei in the vincinity of the doubly magic nucleus $^{78}$Ni are amongst the best candidates to study the evolution of nuclear structure far from stability. In particular the question of whether the N=50 shell gap persists so far away from stability might be studied. The study of the position of the first excited $2^+$ state in even even isotopes can give a first indication on the collectivity of these nuclei. With counting rates as low as a few particles per second the position of the first excited $2^+$ state can be determined [15]. This type of experiments allow for a rather quick "mapping" of a region of interest. As an example, figure 9 shows the evolution of the $2^+$ state energy for neutron-rich isotopes in the N=50 region. Similar studies must be continued toward $^{78}$Ni.

The study of γ rays emitted following spontaneous fission has recently been a precious source of information about the structure of neutron-rich nuclei and about the fission process itself. "in beam" experiments as described in ref [16] and [17] can be very useful. Prompt γ rays following photo-fission can be observed with γ-ray facilities. The multiplicity requirement strongly suppressed events associated with the b decay of fission products. The production of fission fragments by photo-fission with ALTO can be an interesting opportunity to study prompt γ rays in the region of N=50.

Laser spectroscopic studies at ALTO can be used to measure isotope shift and hyperfine structure in atomic spectra. These quantities are used to determine nuclear spins, magnetic dipole and electric spectroscopic quadrupole moments, and to follow the changes in nuclear mean-square charge radius through isotopic sequences. Fig. 10 illustrates the set up that could be used for laser spectroscopic measurements at ALTO. With this set-up, a beam from a tunable laser is made to overlap the beam of ions or atoms produced by ALTO. Measurements could be performed at ALTO for the heavy Ge isotopes, lying above the N=50 shell closure.

To understand stellar evolution and the production of the elements in the universe, extensive model calculations are used to describe and simulate the different processes occurring in the stars. Especially for violent processes like supernovae explosions or X-ray bursts, mainly properties of unstable nuclei are the most important inputs to the models. In neutron-rich stellar environments, the rapid-neutron-capture process produces heavy elements by a sequence of neutron captures and nuclear beta decays. To correctly model this r-process, the model inputs needed are masses of very neutron-rich nuclei, their beta-decay half-lives and their neutron-capture cross sections. However, up to now, these properties are only known for a few isotopes involved in the r process. Neutron-rich fission fragments from ALTO will allow to perform measurements of half-lives and masses for some of the key nuclei. Mass measurements of high precision could be performed at ALTO for ~ 100 new nuclei using the MISTRAL radio-frequency transmission spectrometer from CSNSM at Orsay [18]. MISTRAL is currently installed at the ISOLDE mass separator facility at CERN [19]. The lay-out of MISTRAL is shown in Fig. 11. A production greater than $10^3$ ions per second is needed for the mass measurement program.

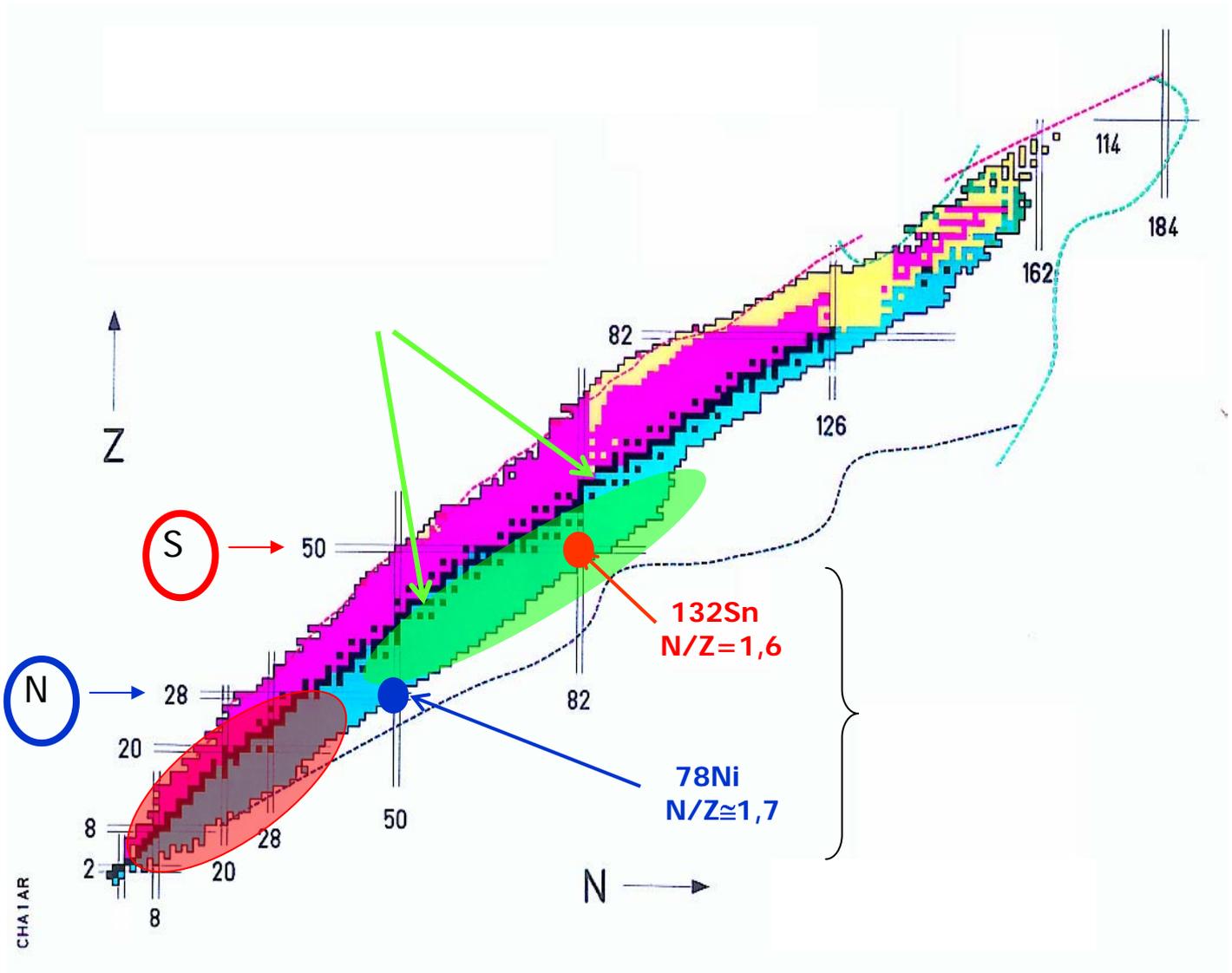

Fig 1  The production of fission fragments (upper light grey part) in comparison with the production for SPIRAL (dark grey part)

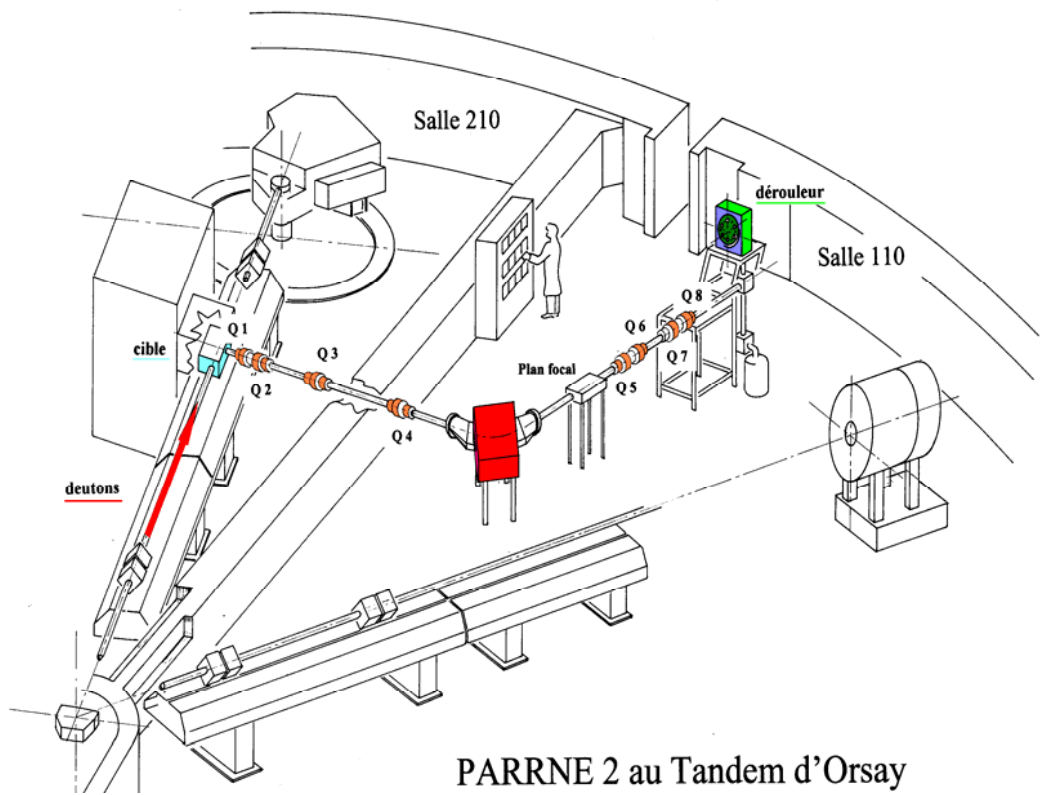

Fig 2 Schematic view of the ISOL device PARRNe 2 at the Tandem of IPN Orsay

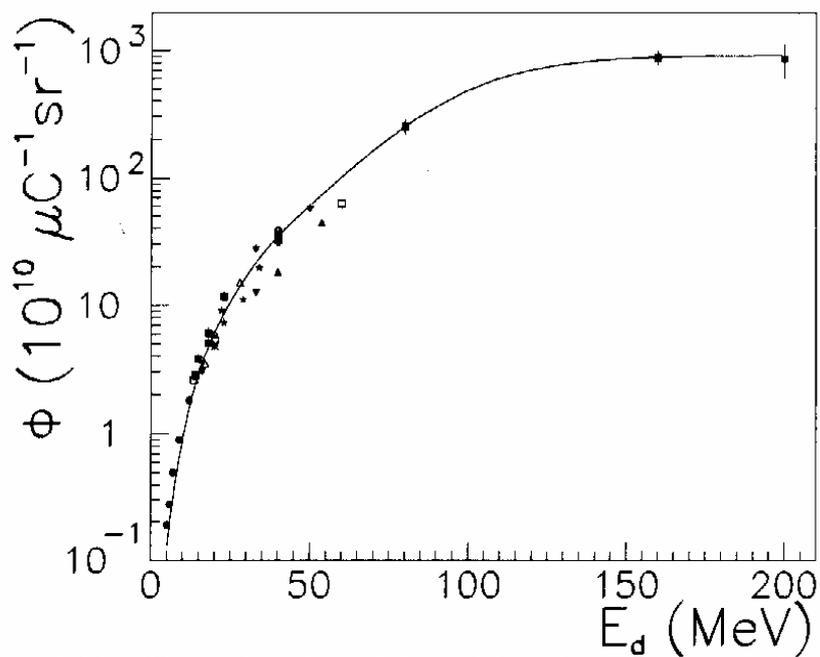

Fig 3  Neutron yield at 0° as a function of the incident deuteron energy  for a Be converter

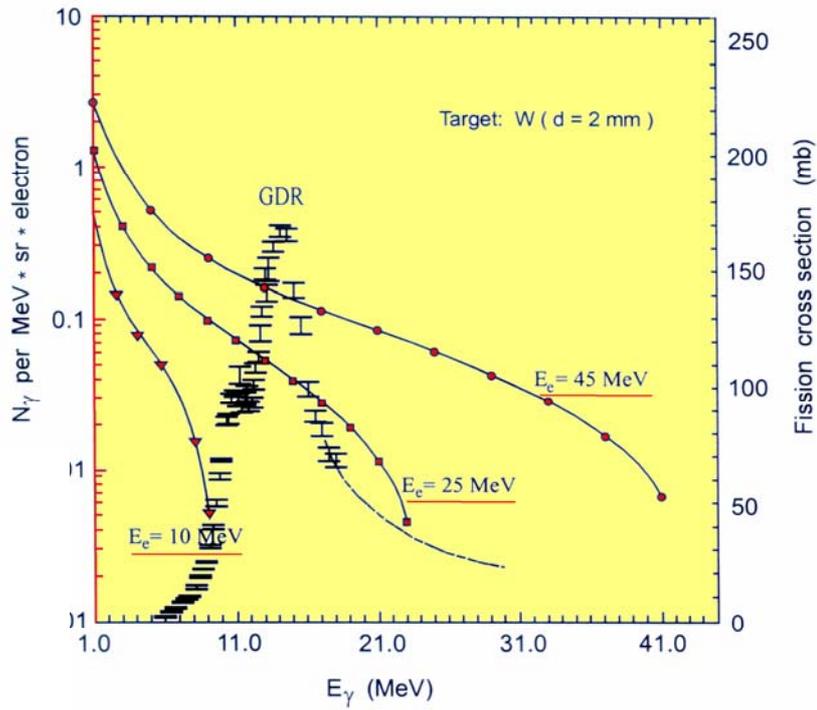

Fig 4  The solid lines is the g-spectrum produced by electrons with various energies. The experimental points correspond to the $^{238}$U photo-fission cross section.

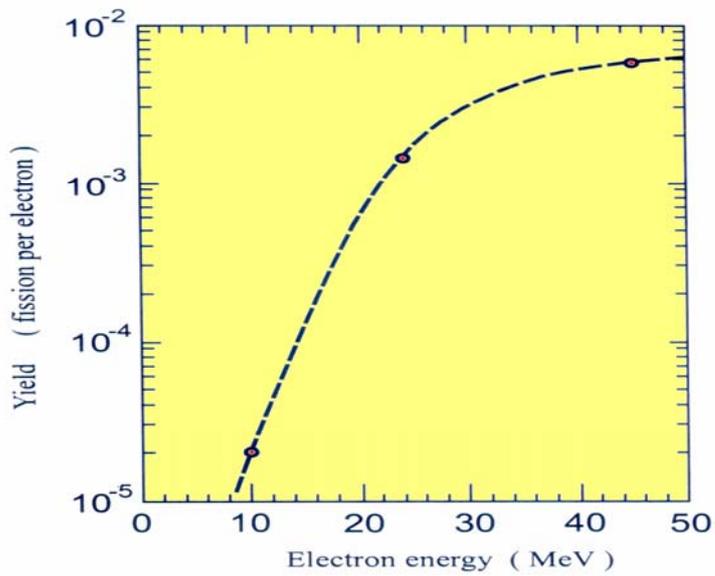

Fig 5  The curve represents the fission yield per electron for $^{238}$U as a function of the electron energy [12]

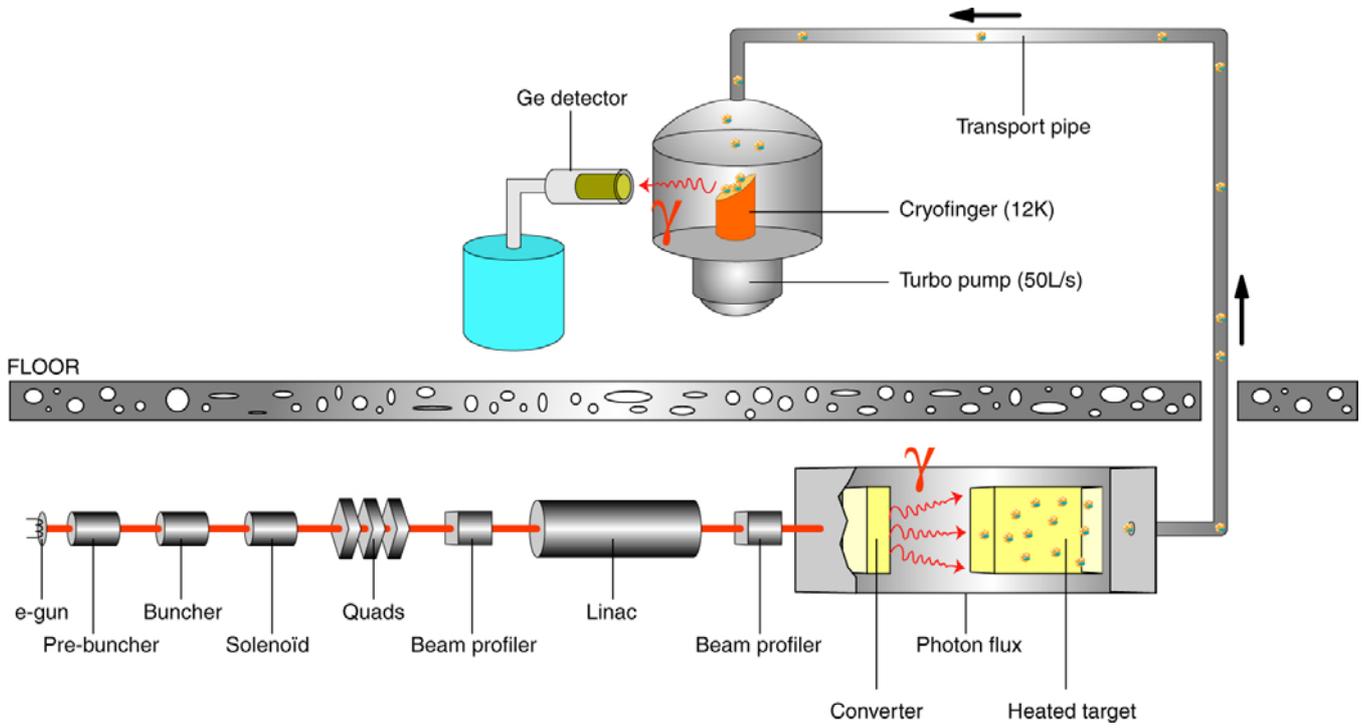

Fig 6 The PARRNe 1 set up at LIL CERN

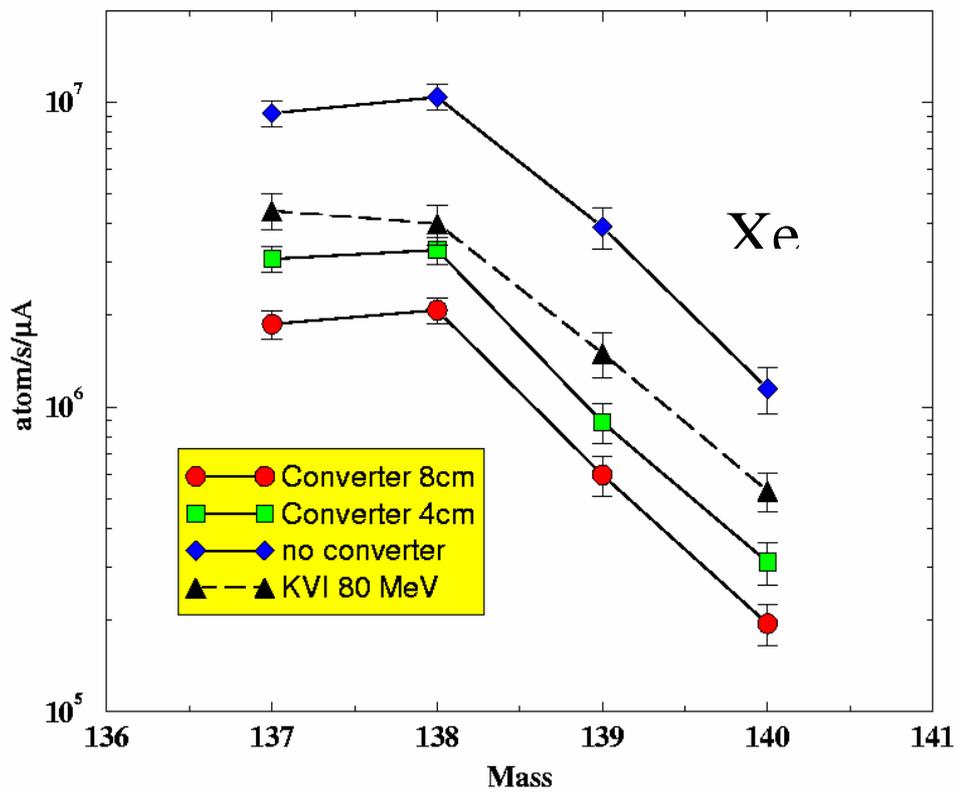

Fig. 7 Results obtained for Xe isotopes using photo-fission

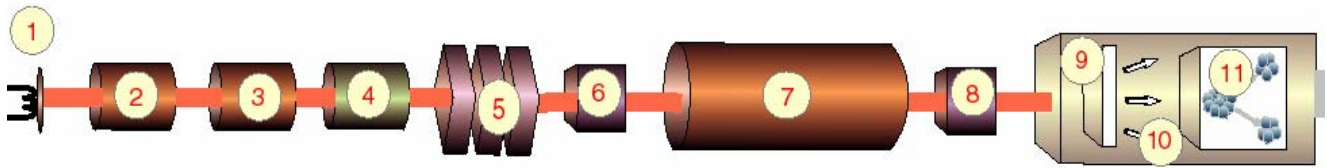

Fig 8  1.e-gun, 2.Pre-buncher, 3.Buncher, 4.Solenoid, 5.Quads, 6.Beam profiler, 7.Acceleration section, 8.Beam profiler, 9.Converter, 10.Photon flux, 11.Heated target

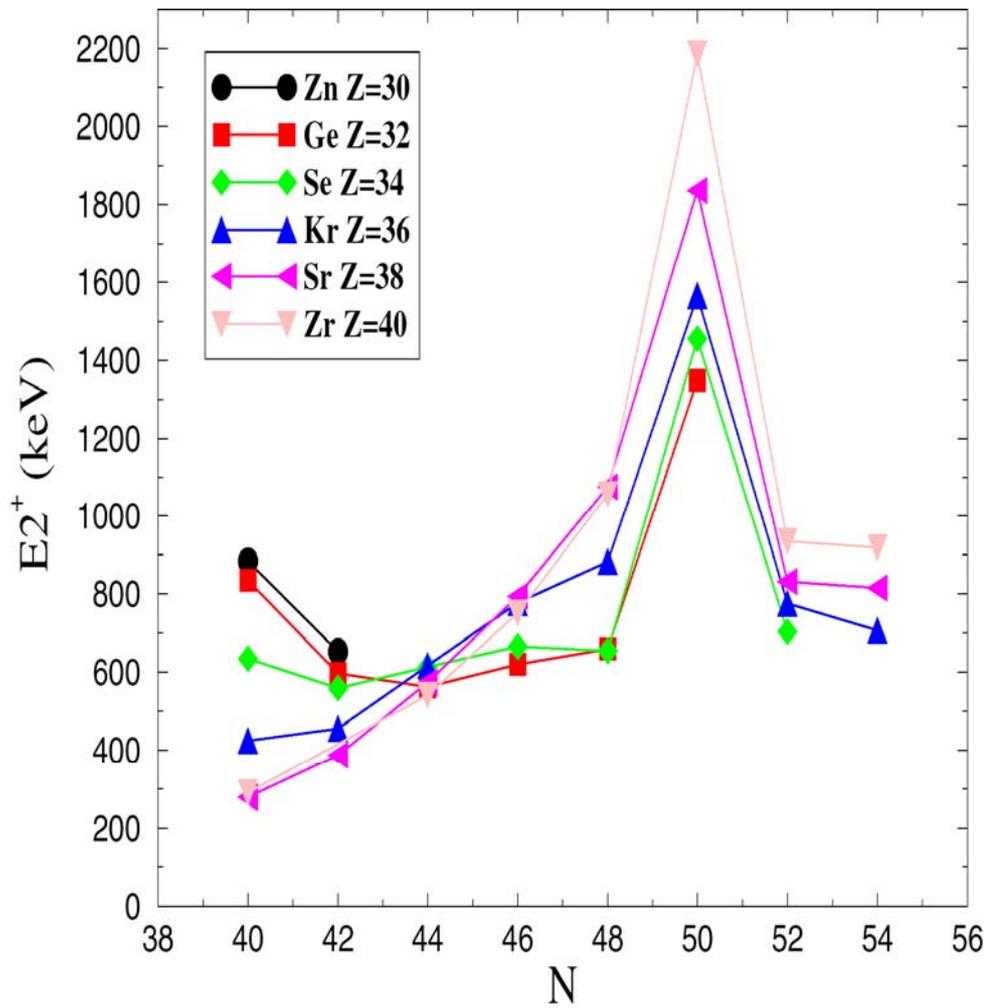

Fig 9 evolution of the $2^+$ energies in the N=50 region

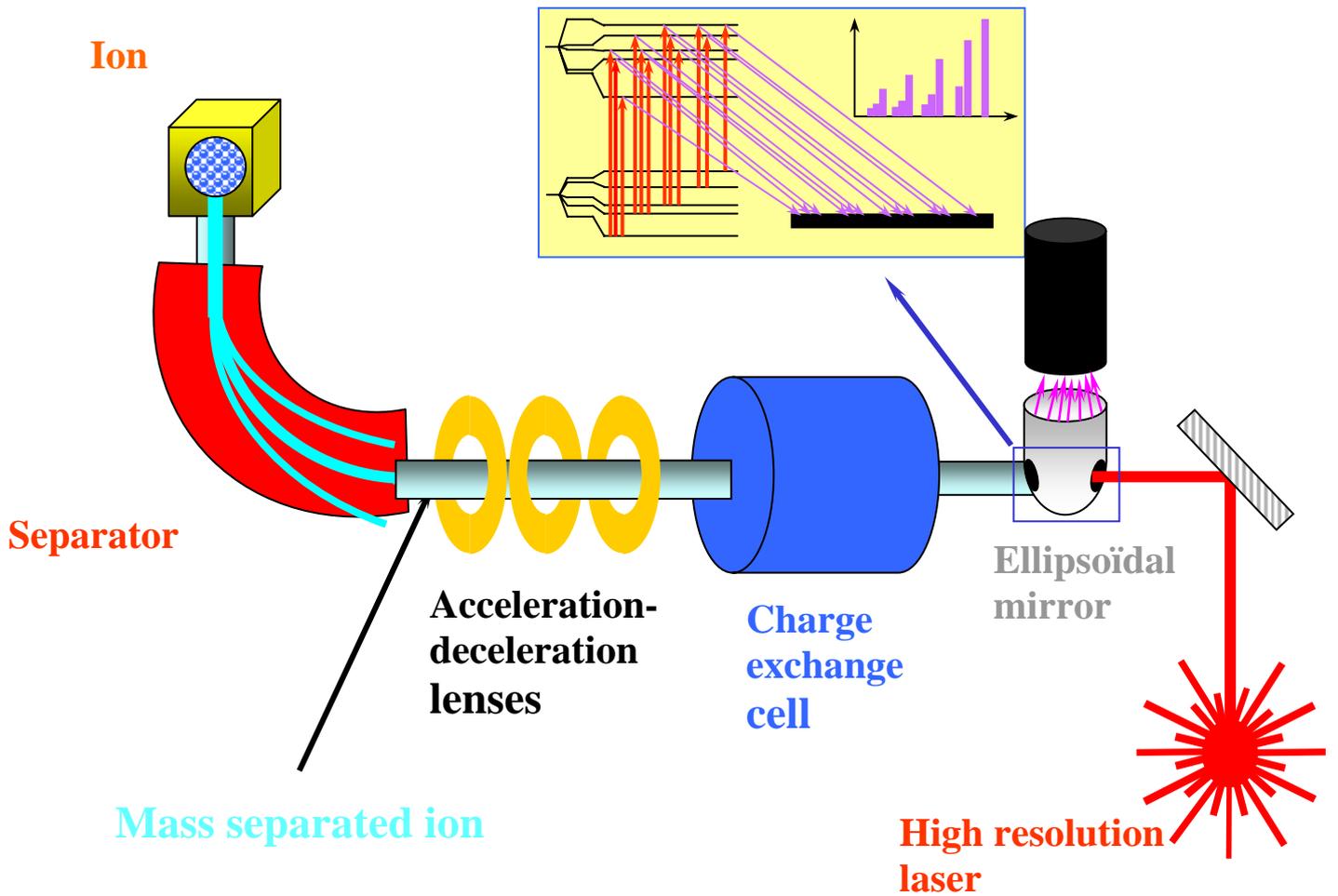

Fig 10   Set up for collinear laser spectroscopy at ALTO

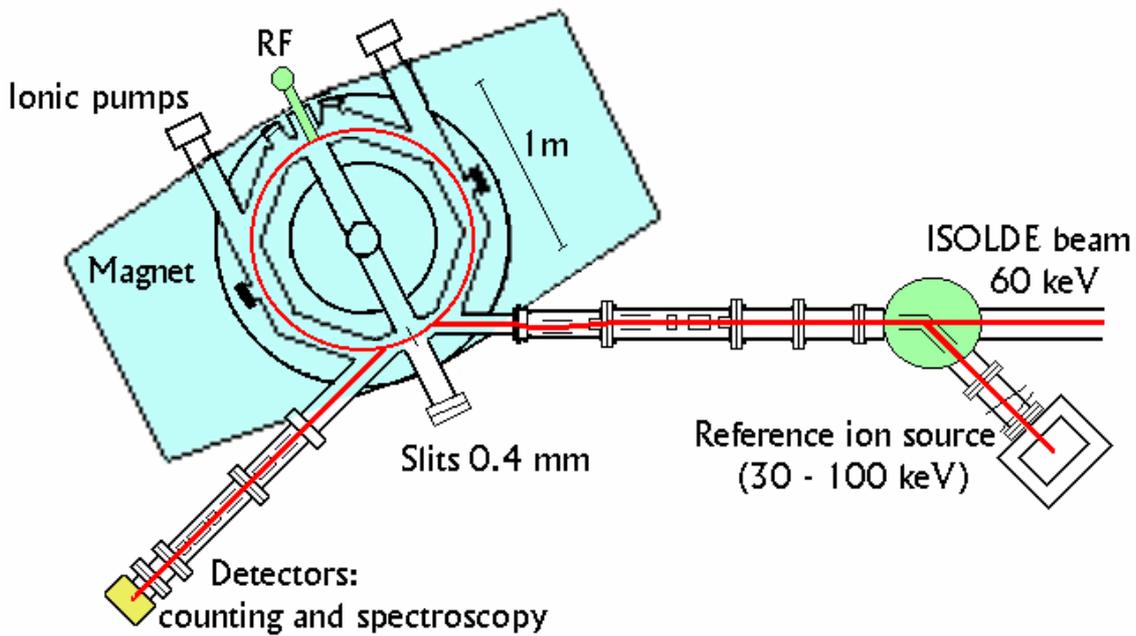

Fig 11   Layout of the MISTRAL spectrometer (overhead view)